\documentclass{ws-ijmpa}
\usepackage[super,compress]{cite}
\usepackage{url}
\usepackage{graphicx}
\usepackage{times}
\usepackage{amsmath}
\usepackage{amsfonts}
\usepackage{epsfig}
\usepackage{dcolumn}
\usepackage{bm}
\usepackage{color}
\usepackage{longtable}
\usepackage{array}
\usepackage{multirow}
\usepackage{axodraw2}
\usepackage{wrapfig}
\usepackage{lipsum}
\usepackage{physics}

\newcommand{\pip}{\pi^+}
\newcommand{\pim}{\pi^-}

\newcommand{\jpsi}{J/\psi}

\newcommand{\ee}{e^+e^-}

\newcommand{\pipi}{\pi^+\pi^-}

\newcommand{\kl}{K_{L}^{0}}

\newcommand{\rt}{\rightarrow}

\newcommand{\beq}{\begin{equation}}
\newcommand{\eeq}{\end{equation}}
\newcommand{\bitm}{\begin{itemize}}
\newcommand{\eitm}{\end{itemize}}

\begin{document}
\markboth{Zhou,~Olsen}{New Window on Matter-Antimatter Differences}
%
\catchline{}{}{}{}{}
%

\title{New Window on Matter-Antimatter Differences from BESIII}

\author{Xiaorong Zhou\footnote{zxrong@mail.ustc.edu.cn}}

\address{State Key Laboratory for Particle Detection and Electronics. Hefei, 230026, China\\
  Department of Modern Physics, University of Science and Technology of
  China, Hefei, 230026, China}

\author{Stephen Lars Olsen\footnote{solsensnu@gmail.com}}

\address{University of Chinese Academy of Science,Beijing 100049, China\\
Institute for Basic Science,  Daejeon 34126, South Korea}

\maketitle


\begin{abstract}
  Recent measurements of $CP$-related comparisons of the weak-interaction parameters that describe the decays
  of the $\Xi^-$ and the $\bar{\Xi}^+$ doubly strange hyperon and antihyperon are reviewed and their significance
  is discussed.
 
\end{abstract}
\ccode{PACS numbers:}

\section{Introduction}
\label{Sec:intro}

The strange particles played an illustrious role in the history of high energy physics and the development
of the Standard Model (SM), the hugely successful theory that describes the properties and mutual interactions
of the most elementary constituents of matter. In fact, the emergence of high energy physics as an
area of research that is distinct from nuclear physics was the construction and operation of the Cosmotron
accelerator at Brookhaven National Laboratory in 1952 and the Bevatron accelerator  at Berkeley in 1955 for the
expressed purpose of producing and studying the ``New Particles,'' {\it i.e.} the mysterious ``V'' particles
that had been discovered in cosmic-ray interactions~\cite{Leprince-ringuet:1944abc,Rochester:1947abc}.
Studies of these ``strange'' new particles that occurred during the next fifteen years under the controlled
laboratory conditions provided by these and follow-up accelerators uncovered many of the fundamental principles
that make up the SM, including:
\begin{description}
\item[Flavor Quantum Numbers:]~``Strangeness'' was the first of the SM's four ``flavors,'' strange, charm,
  bottom,~\&~top;
\item[Particle-antiparticle Mixing:]~$K^0\leftrightarrow\bar{K}^0$-like transitions also occur with charmed and
  bottom particles, and related processes occur with neutrinos;
\item[Parity Violation:]~left-right symmetry violating processes were first seen in  $K$-meson decays
  and are now a defining characteristic of the Weak Interactions;
\item[\boldmath{$CP$} Violation:] the first sign of matter and antimatter differences was the observation of
  $\kl\rt\pipi$ decays;
\item[Flavor mixing:] the suppression of the $\Lambda\rt p e^-\bar{\nu}$ decay rate below that inferred from
 the half-life for $n\rt p e^-\bar{\nu}$ decay was the first sign of flavor-mixing, and
  the precursor of the CKM quark-mixing and PMNS neutrino-mixing matrices;
\item[Quark Model:] the Isospin-Strangeness patterns in the baryon  octet~\&~decuplet and the meson octets
  led to the discovery of $SU(3)$ and fractionally charged quarks;
\item[GIM Mechanism]: the strong suppression of strangeness-changing neutral-current kaon decays led to prediction
  of the  existence of the $c$-quark.
\end{description}

\noindent
Another particle that played a key role in the development of our current understanding of nature is the
$\jpsi$ meson that was discovered nearly simultaneously by experiments at Brookhaven~\cite{Augustin:1974xw}
and SLAC~\cite{Aubert:1974js} in November 1974.  Its appearance at SLAC was especially dramatic, where
it showed up as a spectacular narrow peak in the inclusive cross-section for $\ee\rt{\rm hadrons}$ from its
well established continuum level of $\sigma_{\rm had}\approx 25$~nb to $\sigma_{\rm had}\approx 2500$~nb; a factor of
$\sim 100$ jump contained within a $\sim 0.1\%$~interval of center-of-mass energy. Since the $\jpsi$ is an
$SU(3)$-singlet, its decay strengths to strange particle and non-strange particle final states are similar,
and this, together with its large production cross-section, make the reaction $\ee\rt\jpsi\rt Y\bar{Y}$
($Y=\Lambda,~\Sigma,~\Xi$) a prolific source of strange hyperons. 

After the discovery of the $\jpsi$, studies of strange particles faded from the limelight
as interest in the new charmed and, shortly thereafter, bottom-flavored particles attracted  more and more
of the particle physics community's attention. Although the development of dedicated hyperon beams at
Fermilab and CERN resulted in precise measurements of hyperon magnetic moments~\&~semileptonic decays, and
heroic efforts at the  two labs resulted in the discovery of a direct $CP$-violating component of
$K^0\rt\pi\pi$ decays ({\it i.e.}, a non-zero value of the $CP$ parameter $\varepsilon^{\prime}$), the activities
with the highest priorities were those aimed at measuring the parameters and testing the validity of the SM with
neutrinos, the new heavy-quarks, the intermediate-vector bosons and the Higgs particle.

So far, the SM has prevailed and the main focus of the field has switched to searches for cracks in the theory
that might provide clues as to what physics are not properly addressed by it might look like. In this arena,
precision measurements of strange particle decay properties are as well positioned for finding deviations from
SM predictions as those for any of the newer particles.  Moreover, the hyperons and antihyperons that are produced
in the $\jpsi\rt Y\bar{Y}$ decay process are quantum-correlated, which make them particularly well suited for some
of these searches.  Maybe history will repeat itself and once again it will be strange particles that teach us
the physics that lies beyond the SM.

\section{The role of  $CP$ violations in BSM-physics searches}

In spite of the SM's success at providing precise quantitative explanations for virtually all
experimentally observed phenomena, including $CP$-symmetry violations in the decays of strange and bottom
mesons, it has some obvious deficiencies that indicate that it is not a complete theory. Notable among these
deficiencies is the SM's failure to provide a description of how the current all-matter universe evolved from
what had to have been a matter-antimatter symmetric condition that existed shortly after the Big Bang. The SM
mechanism for $CP$ violation fails to account for the observed baryon asymmetry of the current universe by
some ten orders of magnitude. Therefore it is generally considered that the new, ``Beyond the Standard Model''
(BSM) physics that will supersede the SM must include  at least one new source of $CP$
violation.

Another important issue that confronts virtually all searches for BSM physics is the fact that the Quantum
Chromodynamics sector of the SM, which in principle explains all of strong interaction physics, is, in
practice, woefully inadequate at producing reliable high-precision descriptions of low-energy processes
that involve hadrons. As a result, the sensitivity of most searches for BSM new physics are ultimately
limited by QCD-associated uncertainties. However, thanks to stringent limits on the neutron's electric
dipole moment~\cite{Abel:2020gbr} ($d_n<2\times 10^{-26}e$cm), the validity of $CP$~symmetry in QCD has been
established at the part in $10^{10}$ level~\cite{Baluni:1978rf}. As a result, searches for new physics sources
of $CP$ are relatively immune to QCD-related uncertainties. Searches for anomalous $CP$~violations in the $c$-
and $b$-quark sectors are major motivations for the LHCb experiment~\cite{LHCb:2008vvz} at the CERN Large Hadron
Collider and the Belle~II experiment~\cite{Aushev:2010bq} at the KEK Super~$B$~Factory $\ee$ collider. But, since,
by definition, nothing is known about what the new sources of $CP$ violation might be, there is no
{\it a priori} reason that they won't first show up at measurable levels in the $s$-quark sector. So far,
however, that hasn't happened;  previous searches for anomalous $CP$ violations in
kaon~\cite{Angelopoulos:2003hm} and hyperon~\cite{E756:2000rge} decays have reached the sensitivity limits of
their respective experimental methods and found no unexpected results.

\section{$CP$ studies with quantum-correlated $\Xi\bar{\Xi}$ pairs produced via $\jpsi$ decay}

In this context, we direct the reader to a recent publication from the BESIII experiment at the
Beijing Institute of High Energy Physics BEPCII $\ee$ collider that reports a new approach to the
search for a BSM source of $CP$~violation in the $s$-quark sector. This involves a comparison of the
weak interaction decay parameters of the doubly strange $\Xi^-$ hyperons and $\bar{\Xi}^+$ anti-hyperons
that are produced in quantum mechanically entangled pairs via the
$\ee\rt\jpsi\rt\Xi^-\bar{\Xi}^+$ reaction~\cite{BESIII:2021ypr}.

\begin{figure}[h!]
  \begin{center}
    \includegraphics[width=0.65\textwidth]{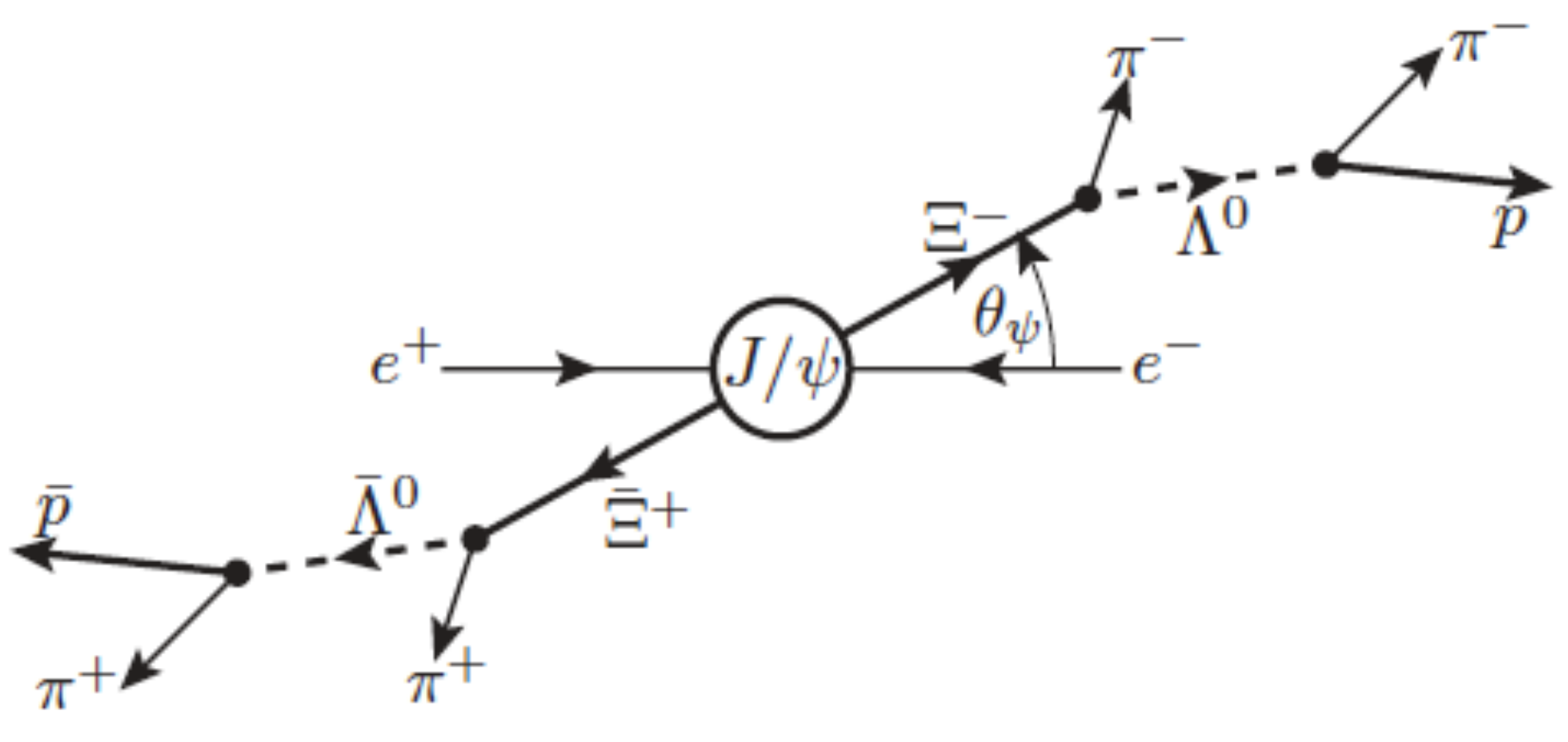}
    \caption{\footnotesize Quantum correlated $\Xi^-\bar{\Xi}^+$ decay chains produced in $\jpsi$ decays. 
    }
    \label{fig:XiXibar}
  \end{center}
\end{figure}

In the BESIII measurement, the $\Xi$ hyperons occur in the  reaction chains shown in Fig.\ref{fig:XiXibar}.
Here the $\jpsi$ mesons are produced (nearly) at rest in the laboratory by counter-circulating $e^+$ and
$e^-$ beams in the BEPCII collider and the $\Xi^-$~\&~$\bar{\Xi}^+$ are produced back-to-back, in a spin-correlated
quantum-entangled state that persists until one of them decays, which usually occurs within a few~centimeters of the
production point. In the selected events, both of the $\Xi$ hyperons decay to a charged pion and a $\Lambda$ hyperon
that, in turn, decays into a charged pion and a proton also within about a few centimeters of the parent $\Xi$ decay
points. The events show up as in the detector as $p\pim$ and $\bar{p}\pip$ decay products of nearly back-to-back
$\Lambda$ and $\bar{\Lambda}$ hyperons that are accompanied by low momentum $\pim$ and $\pip$ tracks. The detected
events have a very clear topology, with highly constrained kinematics and no particle-assignment
ambiguities.\footnote{The laboratory
      momentum ranges of the $p$~\&~$\bar{p}$ and those for the pions have very little overlap.}
Moreover, the background levels are  very small; the signal to background ratio  in the BESIII measurement is 400:1.

The overall process involves five different reactions that are most simply described by considering five different
reference frames. The first is the $\jpsi$ rest frame where, since the spin=1 $\jpsi$ mesons are produced in
$\ee$~annihilation via a single virtual photon, they are spin-aligned along the direction of the $\ee$ beamline
(taken as the $z$ axis). This means that equal numbers of them are produced in the $\ket{J;J_z}=\ket{1;+1}$ and
$\ket{1;-1}$ angular momentum states, but none with $\ket{J;J_z}=\ket{1;0}$. The $\jpsi\rt\Xi^-\bar{\Xi}^+$ decay
matrix element is the sum of two helicity amplitudes. The interference between these two
amplitudes and the absence of any initial-state $J_z=0$ component produces a $\theta_\psi$-dependent production cross
section for $\Xi$~\&~$\bar{\Xi}$ hyperons that are spin-polarized (with polarization vector
$\mathbf{P}_{\Xi}$ ) along the direction perpendicular to the production plane~\cite{Perotti:2018wxm}:
  \begin{eqnarray}
   \label{eqn:jpsi-xixibar-alpha}
   \frac{1}{N}\frac{dN}{d\cos\theta_{\psi}}&=&\frac{3}{4\pi}\frac{1+\alpha_\psi \cos^2\theta_{\psi}}
           {3+\alpha_{\psi}}\\
         \mathcal{P}_{\Xi}&=&\frac{\sqrt{1-\alpha^2_{\psi}}\sin\theta_{\psi}\cos\theta_{\psi}\sin\Delta\Phi}
            {1+\alpha_{\psi}\cos^2\theta_{\psi}},
  \end{eqnarray}
where $\alpha_{\psi}$ is a parity- and $CP$-conserving parameter that is proportional to the product of the two
helicity amplitudes, $\Delta\Phi$ is their relative phase, and $\mathcal{P}_{\Xi}$ is the vector modulus of $\mathbf{P}_{\Xi}$.
 The polarizations of the hyperons are equal,
{\it i.e.}, $\mathbf{P}_{\Xi}=\mathbf{P}_{\bar{\Xi}}$, .

\begin{figure}[h!]
  \begin{center}
    \includegraphics[width=0.95\textwidth]{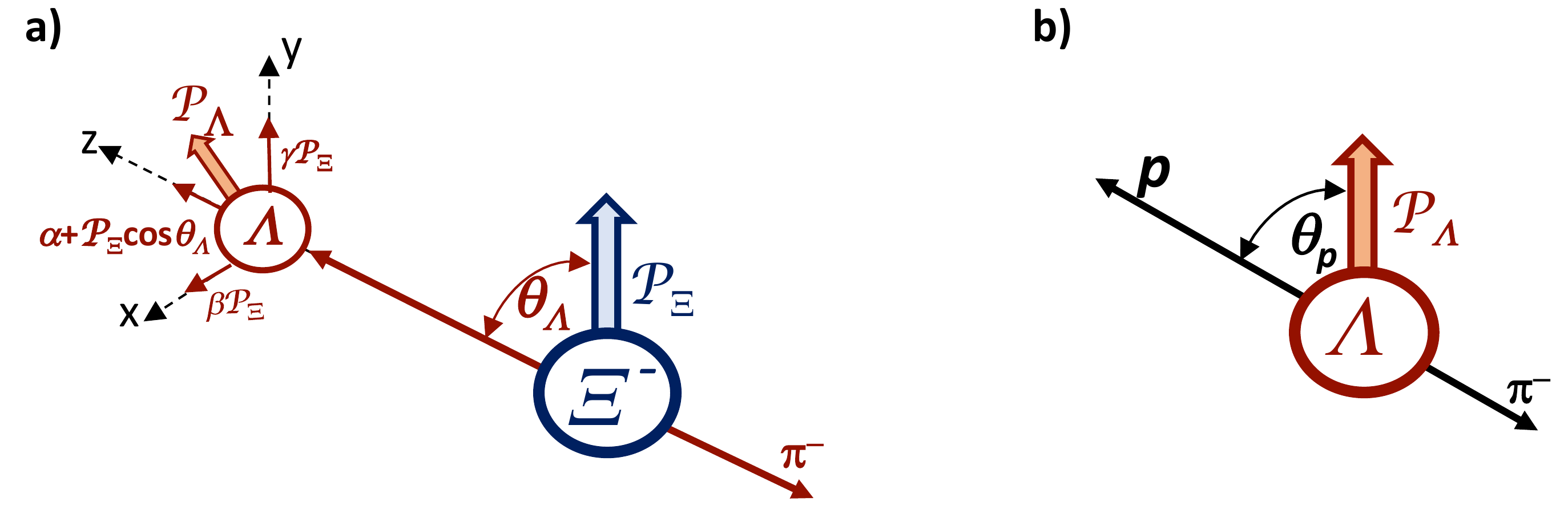}
    \caption{\footnotesize {\bf a)} The decay angle $\theta_{\Lambda}$  and $\Lambda$ polarization vector
      $\mathbf{P}_{\Lambda}$ for polarized $\Xi^-\rt\Lambda\pim$ decays. {\bf b)} The decay angle $\theta_p$
      for polarized $\Lambda\rt p\pim$ decays.}
    \label{fig:Xi_and_Lambda-decays}
  \end{center}
\end{figure}

  The next two reference frames to consider are the $\Xi^-$ and $\bar{\Xi}^+$ rest frames. In these frames
  the decay angle $\theta_\Lambda$ ($\theta_{\bar{\Lambda}}$) relative to $\mathbf{P}_{\Xi}$ 
  direction, and the polarization vectors of the $\Lambda$ daughters, $\mathbf{P}_{\Lambda}$
  $(\mathbf{P}_{\bar{\Lambda}})$, 
  as illustrated in Fig.~\ref{fig:Xi_and_Lambda-decays}a, are distributed according to~\cite{Lee:1957qs}
    \begin{eqnarray}
     &~& \frac{dN}{d\cos\theta_{\Lambda}}\propto 1+\alpha_\Xi\mathcal{P}_{\Xi}\cos\theta_{\Lambda}\\
      \label{eqn:Lambda-pol}
     &~&~~{\bold P}_{\Lambda}=\frac{(\alpha_\Xi+\mathcal{P}_{\Xi}\cos\theta_\Lambda)\mathbf{\hat{z}}
      +{\mathcal P}_{\Xi}\beta_\Xi\mathbf{\hat{x}}
      +{\mathcal P}_{\Xi}\gamma_\Xi\mathbf{\hat{y}}}{1+\alpha_{\Xi}\mathcal{P}_{\Xi}\cos\theta_\Lambda},
    \end{eqnarray}
    where the unit vectors $\mathbf{\hat{x},\hat{y},\hat{z}}$ are oriented as indicated in the figure
    and $\alpha,\beta,\gamma$ are the Lee-Yang decay parameters that are described in the following
    subsection.

\noindent
The last two reference frames are the $\Lambda$ and $\bar{\Lambda}$ rest frames. Here the distribution of
events in $\theta_p$ ($\theta_{\bar{p}}$), the direction of the proton relative to the $\Lambda$ polarization
vector (see Fig.~\ref{fig:Xi_and_Lambda-decays}b), is
    \begin{equation}
      \frac{dN}{d\cos\theta_{p}} \propto 1+\alpha_\Lambda\mathcal{P}_{\Lambda}\cos\theta_{p}.
    \end{equation}
where $\mathcal{P}_{\Lambda}$ is the vector modulus of $\mathbf{P}_{\Lambda}$.
 Since the proton and antiproton polarizations are not measured, only the $\alpha_{\Lambda}$ and
 $\alpha_{\bar{\Lambda}}$ parameters are determined.
 
\subsection{The Lee-Yang $\alpha,\beta,\gamma$ parameters}
The $\Xi\rt\Lambda\pi$ and $\Lambda\rt p\pi$ decays are weak interaction processes that involve both
$S$- and $P$-waves, and are characterized by the (real) Lee-Yang parameters
$\alpha_Y$,~$\beta_Y$~and~$\gamma_Y$~\cite{Lee:1957qs}, where
\begin{equation}
  \alpha_Y = \frac{2\Re(S^*_YP_Y)}{|S_Y|^2+|P_Y|^2},~~~~\beta_Y = \frac{2\Im(S^*_YP_Y)}{|S_Y|^2+|P_Y|^2},
  ~~~~\gamma_Y = \frac{|S_Y|^2-|P_Y|^2}{|S_Y|^2+|P_Y|^2}.
\end{equation}
where $\alpha^2_Y + \beta^2_Y + \gamma^2_Y=1$. The amplitudes $S_Y$ and $P_Y$
are complex and can have two distinct types of phases. One type corresponds the $S$- and $P$-wave $\Lambda\pi$
and $p\pi$ final-state strong interaction phase shifts, $\delta^S_{y\pi}~\&~\delta^P_{y\pi}$, where
$Y\rt y\pi$~($y=\Lambda~{\rm or}~p$); they have the {\it same sign} for particles and antiparticles. If $CP$
is violated, they will have additional so-called {\it weak}-, or $CP$-phases
$\xi^{P}_{Y}$ and $\xi^{S}_{Y}$ that have {\it opposite signs} for particles and antiparticles. Thus, $\alpha_Y$
(hyperon) and $\alpha_{\bar{Y}}$ (anti-hyperon) have the form\cite{Donoghue:1985ww}
\begin{eqnarray}
  \alpha_Y&=&
  \frac{2|S_Y||P_Y|\cos\big((\delta^P_{y\pi}-\delta^S_{y\pi})+(\xi^P_{Y}-\xi^S_{Y})\big)}{|S_Y|^2+|P_Y|^2}\\
\nonumber
  \alpha_{\bar{Y}} &=&
  -\frac{2|S_Y||P_Y|\cos\big((\delta^P_{y\pi}-\delta^S_{y\pi})-(\xi^P_{Y}-\xi^S_{Y})\big)}{|S_Y|^2+|P_Y|^2},
\end{eqnarray}
where the minus sign in the expression for $\alpha_{\bar{Y}}$ comes from the operation of the parity operator
({\it i.e.} the $P$ in $CP$) on the $P$-wave amplitude. Thus, assuming $\xi^P_{Y}-\xi^S_{Y}$ is small,
\begin{equation}
  \label{eqn:ACP-asymmetry}
        {\mathcal A}_{CP}^Y\equiv\frac{\alpha_Y+\alpha_{\bar{Y}}}{\alpha_Y-\alpha_{\bar{Y}}}
        =-\sin(\delta^P_{y\pi}-\delta^S_{y\pi})\sin(\xi^P_{Y}-\xi^S_{Y}) ,
\end{equation}
and, if $CP$ is not conserved, $\xi^P_{Y}\neq\xi^S_{Y}$, and any significant difference between
${\mathcal A}_{CP}^Y$ and zero would unambiguous evidence for a $CP$ violation.

However, it is evident from eqn.~\ref{eqn:ACP-asymmetry} that even if $\xi^P_{Y}-\xi^S_{Y}$ is non-zero,
${\mathcal A}_{CP}^Y$ would be zero if $\delta^P_{y\pi}-\delta^S_{y\pi} =0$. Although it is unlikely that
these strong-interaction phase-shift differences are exactly zero, they are  known to be small -- the
 measured values for the $p\pi$ shifts are are~\cite{Matsinos:2022xll}
$\delta^P_{p\pi}-\delta^S_{p\pi}=-7.2^{\circ}\pm 0.2^{\circ}$ while the theoretical values for
the $\Lambda\pi$ shifts are~\cite{Huang:2017bmx} 
$\delta^P_{\Lambda\pi}-\delta^S_{\Lambda\pi}=8.8^{\circ} \pm 0.2^{\circ}$-- and these will
suppress the values of the measured ${\mathcal A}_{CP}^Y$ asymmetry by an order of magnitude. This is not the case for
measurements of the parameter $\beta_{\Xi}$ that can be determined from the measurement of
$\mathbf{P}_{\Lambda}$ in $\Xi\rt\Lambda\pi$ decays. Here, we apply the above-described calculation to
a different asymmetry parameter ${\mathcal B}_{CP}^{\Xi}$, where 
\begin{equation}
  \label{eqn:BCP-asymmetry}
        {\mathcal B}_{CP}^{\Xi}\equiv\frac{\beta_{\Xi}+\beta_{\bar{\Xi}}}{\alpha_{\Xi}-\alpha_{\bar{\Xi}}}
    = \tan(\xi^P_{\Xi}-\xi^S_{\Xi})\approx \xi^P_{\Xi}-\xi^S_{\Xi}.
\end{equation}
(Here $\beta_{\Xi}+\beta_{\bar{\Xi}}$ is divided by $\alpha_{\Xi}-\alpha_{\bar{\Xi}}$ and not
         $\beta_{\Xi}-\beta_{\bar{\Xi}}$ in order to avoid the perils of a near-zero denominator.)
In this case the measured asymmetry doesn't suffer any dilution by small strong-interaction
phase-shifts.  Since the $\beta_{\Xi}$ and $\gamma_{\Xi}$ parameters are not independent, BESIII reports
measurements of $\phi_{\Xi}$ (and $\phi_{\bar{\Xi}}$) that are defined by the relations~\cite{Donoghue:1986hh}
\begin{equation}
  \beta_{\Xi}=\sqrt{1-\alpha^2_{\Xi}}\sin\phi_{\Xi}~~{\rm and}
  ~~\gamma_{\Xi}=\sqrt{1-\alpha^2_{\Xi}}\cos\phi_{\Xi}.
\end{equation}
In terms of  $\Delta\phi_{CP}^{\Xi}\equiv {\textstyle \frac{1}{2}} (\phi_{\Xi}+\phi_{\bar{\Xi}})$ and
$\langle\phi_{CP}^{\Xi}\rangle\equiv {\textstyle \frac{1}{2}} (\phi_{\Xi}-\phi_{\bar{\Xi}})$:
\begin{eqnarray}
  \label{eqn:Deltaphi_CP}~
  \xi^P_{\Xi}-\xi^S_{\Xi}&=&\frac{\sqrt{1-\alpha^2_{\Xi}}}{\alpha_{\Xi}}\Delta\phi_{CP}^{\Xi}\\
  \nonumber
  \delta^{P}_{\Lambda\pi}-\delta^{S}_{\Lambda\pi}&=&
        \frac{\sqrt{1-\alpha^2_{\Xi}}}{\alpha_{\Xi}}\langle\phi_{CP}^{\Xi}\rangle.
\end{eqnarray}

The only direct $CP$-violation ever seen in the strange particle sector is the kaon parameter
$\varepsilon^{\prime}$ that was measured as a  small, non-zero effect in $\kl\rt\pi\pi$ decay
by the NA48 experiment at CERN~\cite{NA48:2001bct} and the KTeV experiment at Fermilab~\cite{KTeV:1999kad}.
This parameter quantifies the $CP$-violating phase difference between the Isospin=2 and Isospin=0,
parity-violating $S$-wave amplitudes in $K^0(\bar{K}^0)\rt\pi\pi$ decays and is suppressed in magnitude
by a nominal factor of $1/22.5$ by the weak-interaction's $\Delta I=1/2$~rule. The $CP$-phase of the ($S$-wave)
amplitude $K\rt\pi\pi$ that is deduced from the measured value of $\varepsilon^{\prime}$ and adjusted
to account for the $\Delta I=1/2$ rule suppression\footnote{This is determined from
  $\xi^S_{K\rt\pi\pi}\approx 22.5\times\sqrt{2}\Re(\varepsilon^{\prime}/\varepsilon)\times\varepsilon$,
  where by definition $|\Im A^{I=2}_{K\rt\pi\pi}|=\sqrt{2}|\Re(\varepsilon^{\prime}|/\varepsilon)|$, of which
  $\Re(\varepsilon^{\prime}/\varepsilon)=(1.66\pm 0.23)\times 10^{-3}$ is what is measured,
  $\varepsilon=(2.228\pm 0.011)\times 10^{-3}$ is the neutral kaon mass-matrix $CP$-violating
  parameter, and the factor~of~22.5 accounts for the $\Delta I=1/2$ rule suppression.}
is $\xi^S_{K\rt\pi\pi}\approx 0.005^{\circ}$, and is within errors of the SM prediction~\cite{Buras:2020wyv}
$\xi^S_{K\rt\pi\pi}(SM)\approx 0.004^{\circ}$. Thismeasured  magnitude of the $CP$-phase in
(parity-violating) $K\rt\pi\pi$ decay and its agreement with theory has been
translated~\cite{Tandean:2002vy,He:1999bv} into stringent limits on contributions from BSM new physics
sources to $\xi^S_{\Xi}$  (and $\xi^S_{\Lambda}$), and this implies that if any $CP$~violation is
seen in hyperon decay at the current levels of experimental sensitivity, it would most likely 
correspond to enhancements in $\xi_\Xi^P$, {\it i.e.} to the parity-conserving $P$-wave amplitude.

\subsection{Advantages of the $\ee\rt\jpsi\rt\Xi^-\bar{\Xi}^+$ environment}

\noindent
In the BESIII measurement, which uses fully reconstructed events of the above-described decay
chains, the $\Xi^-$ and $\bar{\Xi}^+$ are produced in equal numbers, with equal polarizations and  momenta, and
under the same experimental conditions.  Moreover, the decay products are detected in a low mass Helium-filled
tracking volume that minimizes the main irreducible systematic differences between the $\Xi$ and $\bar{\Xi}$
decay chains, which are the different nuclear interaction probabilities of $p/\bar{p}$ and
$\Lambda /\bar{\Lambda}$ daughter particles in the material of the detector. In BESIII the influence of these
differences are reduced to small second-order effects.

\begin{figure}[h!]
   \begin{center}
    \includegraphics[width=0.95\textwidth]{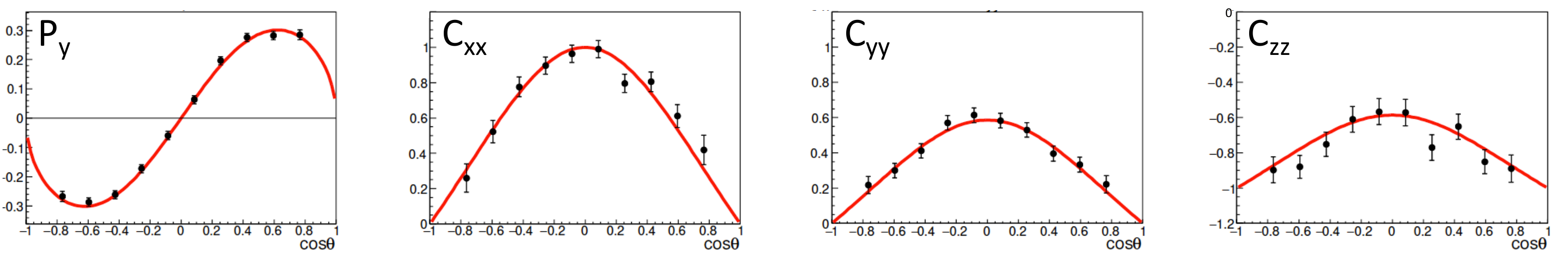}
    \caption{\footnotesize $\Xi$ polarization $\mathbf{P}_y$ and $\Xi$-$\bar{\Xi}$ spin correlations  $C_{xx}$,
     $C_{yy}$ and $C_{zz}$ in quantum entangled $\jpsi\rt\Xi\bar{\Xi}$ from BESIII~\cite{BESIII:2021ypr}.}
    \label{fig:Xi-polarization}
  \end{center}
\end{figure}

A second important feature of this reaction is the BESIII observation of substantial
$\Xi$-hyperon polarization in the $\jpsi\rt\Xi^-\bar{\Xi}^+$ decay process~\cite{BESIII:2021ypr} as shown
in Fig.~\ref{fig:Xi-polarization}, where the rms average is
$\langle \mathbf{P}_\Xi\rangle_{\rm rms}=22.3\%$. This is important because the $\alpha_{\Xi}$ and
$\phi_{\Xi}$ measurement sensitivities are directly proportional to the magnitude of  $\mathbf{P}_{\Xi}$. Note
that according to eqn.~\ref{eqn:Lambda-pol}, the polarization of the daughter $\Lambda$ in $\mathbf{\hat{z}}$ direction is
$\mathcal{P}_{\Lambda} \propto \alpha_{\Xi}+\mathcal{P}_{\Xi}\cos\theta_{\Lambda}$, and, since $\alpha_{\Xi}\approx -0.37$, the
$\Lambda$ hyperons that are produced in $\Xi^-\rt\Lambda\pim$ decays are substantially polarized even at
 production angles where $\mathbf{P}_{\Xi}=0$. 

\noindent

A third feature of this reaction is that the $\Xi~\&~\bar{\Xi}$ are quantum entangled,
with strongly correlated $x$,~$y$~\&~$z$ spin components. These correlations are characterized by the
$\theta_{\psi}$-dependent quantities $C_{xx}$, $C_{yy}$~\&~$C_{zz}$ shown in Fig.~\ref{fig:Xi-polarization}, where
it can be seen that they have similar magnitudes as the linear polarization $\mathbf{P}_y$.  However, unlike
the $\mathbf{P}_y$ terms, which are multiplied by functions of either the $\Xi$ or the $\bar{\Xi}$ decay angles
but not both, the $C_{jj}$ terms are multiplied by functions of both the $\Xi$ and $\bar{\Xi}$ decay angles. In
inclusive (so-called
``single-tag'') measurements, where the decay angles of {\it only} the $\Xi$ or {\it only} the $\bar{\Xi}$ are
considered, the effects of the $C_{ij}$ terms average out to zero and only the $\mathbf{P}_y$ terms
survive.~\cite{Salone:2022lpt} In contrast, in exclusive (``double-tag'') measurements, where the decay angles
of {\it both} the $\Xi$ and the $\bar{\Xi}$ and their correlations, are considered event-by-event, the non-zero
$C_{ij}$ terms have the effect of increasing the {\it effective} polarization and thereby enhance the experimental
sensitivity. This is a big effect. In the BESIII measurement, these correlations enhance the statistical
sensitivity of the $\Delta^{\Xi}_{CP}$ measurement over that for polarized, but unentangled, $\Xi$-$\bar{\Xi}$
pairs, by a factor of $\sim 2.5$; this is equivalent to a six-fold increase in the number of
events~\cite{Salone:2022lpt}. This is the huge benefit of having quantum entangled $\Xi\bar{\Xi}$ pairs.

The fourth big advantage of this reaction is that the $\Lambda$ and $\bar{\Lambda}$ polarizations are determined
event-by-event and, thus, $\phi_{\Xi}$~and~$\phi_{\bar{\Xi}}$ can be measured. This is important because, according to
eqn.~\ref{eqn:Deltaphi_CP}, $\xi^P_{\Xi}-\xi^S_{\Xi}$ can be directly determined from $\phi_{\Xi}$~and~$\phi_{\bar{\Xi}}$
without any dilution from the small $\Lambda\pi$ strong interaction phase shifts.  This increases the experimental
sensitivity to $CP$ phases over that for measurements based solely on ${\mathcal A}^Y_{CP}$ measurements by an
order of magnitude.

\subsection{Analysis and results}

The simplicity of the event topology belies a rather intricate analysis. Nine angular
measurements are needed to completely specify the kinematics of the event: the production angle,
$\theta_{\psi}$, and a pair of angles, $\theta_Y~\&~\varphi_Y$ for each of the four hyperon decay vertices.
The function that describes these highly correlated angular distributions is characterized by eight
parameters: $\alpha_{\psi}$~\&~$\Delta\Phi$ for the $\jpsi\rt\Xi^-\bar{\Xi}^+$ decay vertex, two parameters
for each $\Xi$ decay and one parameter for each $\Lambda$ decay (in which the proton polarization is not
measured). The measured quantities are expressed as a nine-component vector ${\boldsymbol \xi}$ and the
angular distribution by an eight-component vector ${\boldsymbol \omega}$:
\begin{eqnarray}
   \label{eqn:W}
  {\boldsymbol \xi} &=&
  (\theta_{\psi},\theta_{\Lambda},\varphi_{\Lambda},\theta_{\bar{\Lambda}},\varphi_{\bar{\Lambda}},\theta_{p},\varphi_{p},
    \theta_{\bar{p}},\varphi_{\bar{p}})\\
    {\boldsymbol \omega} &=&
  (\alpha_{\psi},\Delta\Phi,\alpha_{\Xi},\phi_{\Xi},\alpha_{\bar{\Xi}},\phi_{\bar{\Xi}},\alpha_{\Lambda},
    \alpha_{\bar{\Lambda}}).
\end{eqnarray}
Perotti~{\it et al.}~\cite{Perotti:2018wxm} provide a modularized expression for the nine dimensional probability
distribution that is in a convenient form for symbolic computing:\footnote{For inclusive single-tag
  $\Xi$ measurements the probability distribution is
  \begin{equation}
  {\mathcal W}({\boldsymbol \xi}:{\boldsymbol \omega})=\sum^{3}_{\mu=0}C_{\mu 0}
  \bigg(\sum^3_{\mu'=0}a^{\Xi}_{\mu\mu'}a^{\Lambda}_{\mu'0}\bigg),
  \end{equation}
  and only the first column of the $C_{\mu\nu}$ matrix, which contains $\mathbf{P}_{y}$ but
  no $C_{jj}$ terms, contributes.}
\begin{equation}
  \label{eqn:W}
  {\mathcal W}({\boldsymbol \xi}:{\boldsymbol \omega})=\sum^{3}_{\mu\nu=0}C_{\mu\nu}
  \bigg(\sum^3_{\mu'=0}a^{\Xi}_{\mu\mu'}a^{\Lambda}_{\mu'0}\bigg)
           \bigg(\sum^3_{\nu'=0}a^{\bar{\Xi}}_{\nu\nu'}a^{\bar{\Lambda}}_{\nu'0}\bigg),
\end{equation}
where $C_{\mu\nu}$ is  a $4\times 4$ matrix that contains the $\theta_{\psi}$-dependent $\Xi~\&~\bar{\Xi}$
polarizations and spin correlations, and each $a^Y_{\alpha\alpha^{\prime}}$ represents a $4\times 4$ matrix with
elements that contain expressions that involve  $\alpha_Y~\&~\phi_Y$ parameters and Wigner D-functions of the
$\theta_Y,\varphi_Y$ angular variables. The sums in eqn.~\ref{eqn:W} involve a total of 256 expressions, of
which 100 are non-zero.

The BESIII measurements~\cite{BESIII:2021ypr} are based on an analysis of 73k fully reconstructed events found in
a data sample that contains a total of 1.3B~$\jpsi$ decays to all modes. The results include the world's most
precise measurement of $\alpha_{\Xi}$ and the first ever measurements of $\alpha_{\bar{\Xi}}$, $\phi_{\bar{\Xi}}$ and
${\mathcal A}^{\Xi}_{CP}$. The precision level of the $\alpha_{\Xi}$ and ${\mathcal A}^{\Xi}_{CP}$ measurements based
on 73k~$\jpsi\rt\Xi^-\bar{\Xi}^+$ events is very nearly the same as that for previous BESIII measurements of
$\alpha_{\Lambda}$ and ${\mathcal A}^{\Lambda}_{CP}$ that were based on a sample of 420k~$\jpsi\rt\Lambda\bar{\Lambda}$
events~\cite{BESIII:2018cnd}. This reflects the benefits of the event-by-event baryon-antibaryon spin correlations
that are mentioned above. Moreover, the 73k~$\Xi^-\bar{\Xi}^+$ events provided independent measurements of
$\alpha_{\Lambda}$,~$\alpha_{\bar{\Lambda}}$~and~${\mathcal A}^{\Lambda}_{CP}$ with similar precision as those from the
earlier higher-statistics BESIII measurements. This is because the rms average polarization of the $\Lambda$ and
$\bar{\Lambda}$ hyperons produced via $\Xi$ decays is about twice that of those produced directly
in $\jpsi\rt\Lambda\bar{\Lambda}$ decays. These new $\alpha_{\Lambda}$ and $\alpha_{\bar{\Lambda}}$ measurements
provide independent confirmation of the earlier BESIII results that are $5\sigma$ higher that their (45 year old)
pre-2018 world average values.

Of particular note are the BESIII measurements of $\phi_{\Xi}$ and $\phi_{\bar{\Xi}}$:
\begin{equation}
 \phi_{\Xi}=0.6^{\circ}\pm 1.1^{\circ}\pm 0.5^{\circ}~~~{\rm and}~~~\phi_{\bar{\Xi}}=-1.2^{\circ}\pm 1.1^{\circ}\pm 0.4^{\circ},
\end{equation}
that, according to eqns.~\ref{eqn:Deltaphi_CP}, translate into
\begin{eqnarray}
  \label{eqn:Delta-phi_CP-results}
    \xi^P_{\Xi}-\xi^S_{\Xi} & = & ~0.7^{\circ}\pm 2.0^{\circ}~~\in \{-2.6^{\circ},+4.0^\circ \}~~{\rm (90\%~C.L.)}\\
  \label{eqn:strong-phase-limit}
  \delta^{P}_{\Lambda\pi}-\delta^{S}_{\Lambda\pi}&=&
                   -2.3^{\circ}\pm 2.1^{\circ}~~\in \{-5.8^{\circ},+1.2^\circ \}~~{\rm (90\%~C.L.)}
\end{eqnarray}
where the errors are the quadratic sums of statistical and systematic errors, with correlations between
the $\phi_{\Xi}$ and $\phi_{\bar{\Xi}}$ taken into account, and our estimates of the 90\% confidence level
allowed intervals are indicated. These limits teach us a few important things:
\begin{itemize}
\item  The eqn.~\ref{eqn:Delta-phi_CP-results} $CP$-phase limit interval, which is
  based on an analysis of the 73k $\Xi^-\bar{\Xi}^+$ events found in a sample of 1.3B $\jpsi$ decays, is
  quite similar to the one based on a recently submitted BESIII measurement of the $\Lambda$ asymmetry
  parameter with 3.2M~fully reconstructed $\Lambda\bar{\Lambda}$ events that were found in a sample of
  10B $\jpsi$ decays~\cite{BESIII:2022qax}:
  \begin{eqnarray}
  {\mathcal A}^{\Lambda}_{CP} &=& -0.0025\pm 0.0047\\
  \nonumber
    \xi^P_{\Lambda}-\xi^S_{\Lambda} &=& (-1.1 \pm 2.1)^{\circ}~~\in \{-4.5^{\circ},+2.1^\circ \}~~{\rm (90\%~C.L.)}.
   \end{eqnarray}
  The $\Delta\phi^\Xi_{CP}$-measurements achieve the same sensitivity as those based on ${\mathcal A}^\Lambda_{CP}$
  with a factor of 40~fewer events.   \color{black}
\item There is still considerable room left for the possible influence of new physics in non-leptonic
  hyperon decay processes before the level of SM effects start to show up. The SM values for the $\xi^P_{\Xi}$
  and $\xi^S_{\Xi}$ $CP$-violating phases have nearly equal but opposite-sign values with the net
  result~\cite{Tandean:2002vy}
    \begin{equation}
  (\xi^P_{\Xi}-\xi^S_{\Xi})_{\rm SM}= -(1.4\pm 1.2)A^2\lambda^5\eta=-0.01^{\circ}\pm 0.01^{\circ},
  \end{equation}
    where $A^2\lambda^5\eta$ is the imaginary part of the $V^*_{td}V^{~}_{ts}$ CKM matrix element product in
    the  Wolfenstein parametrization~\cite{Wolfenstein:1983yz}.
\item Since BESIII's measured value for the
  $\delta^{P}_{\Lambda\pi}-\delta^{P}_{\Lambda\pi}$ strong-phase difference is $<0.1$~rad
  (eqn.~\ref{eqn:strong-phase-limit}), the limit on $\xi^P_{\Xi}-\xi^S_{\Xi}$ that can be inferred
  from ${\mathcal A}_{CP}^{\Xi}$ is almost an order of magnitude less stringent than that derived from
  the $\Delta\phi_{CP}^{\Xi}$ measurement.
\end{itemize}

The only previous $CP$-related experimental result that involved $\Xi$ hyperons is
a HyperCP measurement~\cite{HyperCP:2004zvh}  of
${\mathcal A}_{CP}^{\Xi\Lambda}\equiv (\alpha_{\Xi}\alpha_{\Lambda}-\alpha_{\bar{\Xi}}\alpha_{\bar{\Lambda}})/
(\alpha_{\Xi}\alpha_{\Lambda}+\alpha_{\bar{\Xi}}\alpha_{\bar{\Lambda}})=(0.0\pm 0.7)\times 10^{-3}$,
using $\Xi^-\rt\pim\Lambda(\pim p)$ decays in a $\Xi^-$ beam with an average polarization of
$\langle{\mathcal P}_{\Xi}\rangle\approx 0.035$, and corresponding measurements with a $\bar{\Xi}^+$ beam.
This is a mixture of $\Xi$ and $\Lambda$ asymmetry parameters that cannot be directly compared to the BESIII
individual ${\mathcal A}_{CP}^{\Xi}$ and ${\mathcal A}_{CP}^{\Lambda}$ measurements, and are also subject to the
same provisos about dilution by small strong-interaction phase shifts. HyperCP also used their $\Xi^-$ beam
data to make the first determination of $\phi_{\Xi}=-2.4^{\circ}\pm 0.8^{\circ}$, which differs from the
BESIII measurement by about 2.5 standard deviations~\cite{HyperCP:2004not}.

\section{Future prospects}

The BESIII results described here are based on a 1.3B $\jpsi$ event sample. At the current time, BESIII has
accumulated a total of 10B $\jpsi$ events that are still being analyzed. Thus, we can expect an improvement
in the experimental sensitivity for $\xi^P_{\Xi}-\xi^S_{\Xi}$ to the $\sim 1^{\circ}$ level in the not so distant
future. Moreover, the above-described analysis applies equally well to the similar number of
$\jpsi\rt\Xi^0\bar{\Xi}^0$ events that exist in the BESIII 10B $\jpsi$-event sample that, when analyzed, should
have a similar level of $CP$ sensitivity.

However, this level of sensitivity is still two orders of magnitude above the SM expectations that are described
above, and new physics may be lurking anywhere in between. Substantial improvements on the existing BESIII limit
and those expected in the near future using the above-described techniques would require data samples that contain
 trillions of $\jpsi$ events.

In their most recent data-taking run at the peak energy of the $\jpsi$ ($E_{\rm cm}=3.096$~GeV), the BESIII group
collected about 1B $\jpsi$-events/month. Although this was a remarkable achievement, that rate is well
below what would be needed to accumulate trillions of $\jpsi$ events. The BEPCII collider design was optimized for
studies of charmed mesons that are produced in the process $\ee\rt\psi(3770)\rt D\bar{D}$ at $E_{\rm cm}=3.773$~GeV
with a peak luminosity of $10^{33}$cm$^{-2}$s$^{-1}$, which, at that time ($circa~2005$), was the state-of-the-art
for $\ee$ colliders at that energy. In BEPCII, the $e^+$ and $e^-$ beam bunches are about 1~cm in length and cross
at small angles ($\pm 0.6^{\circ}$) in the horizontal plane. In this case, the luminosity suffers from the so-called
``hour-glass effect'' that restricts how tightly the beams can be compressed at the collision point; this limits
the maximum luminosity to be at the $10^{33}$cm$^{-2}$s$^{-1}$ level. During the two decades that followed the time
that BEPCII was designed, there have been a number of innovations in $\ee$ collider technology~\cite{Raimondi:2006gq}
that would, in principle, allow for further reductions of the beam sizes at the interaction point that could evade
hour-glass effect limitations and potentially  enhance the luminosity by as many as two orders of magnitude. These
include: 
\begin{description}
\item[Nanobeam collisions:]
 a scheme in which very flat beams cross at large angles ($\sim \pm 1.7^{\circ}$)
  in the horizontal plane.
\item[Crab-waist collision scheme:] Where the focus point ({\it i.e.}, the ``waist'') for particles on one of
  the horizontal sides of a beam is advanced, while that on the other side of the same beam is
  retarded so that, in spite of the large crossing angle the waists of all of the particles in both of the
  two beams are synchronized, thereby avoiding the hourglass effect.
\end{description}
Although the incorporation of these schemes in the DAFNE~\cite{Zobov:2016sxm} and the
SuperKEKB~\cite{Onishi:2021ffw} have provided proof-of-principle for these luminosity-enhancement schemes,
they have not yet been translated  into the hoped for order-of-magnitude luminosity improvement. The full
exploitation of these schemes, or variations of them, in a working $\ee$ collider is a high priority issue
in the world's accelerator physics community;\footnote{Many of the issues that limit the luminosity
  in low-energy ``flavor-factory'' colliders also apply to proposed high energy colliders, such
  as the FCC-ee~\cite{Agapov:2022bhm} at CERN and the CEPC~\cite{Geng:2015upa} in China.}
the critical factors that are limiting the luminosity have been identified and steady progress is being
made~to solve them~\cite{Zhou:2021abc}. This progress and the potential payoff in higher luminosity have
inspired proposals in China~\cite{Lan:2021tdy} and in Russia~\cite{Levichev:2008zz} for futuristic
``Super~$\tau$-charm~factories''with luminosity goals that are $\sim 50\times$ that of BEPCII. With either,
or both, of these facilities, the accumulation of trillion (or more) $\jpsi$-event data samples would be
feasible, and an order magnitude improvement on the BESIII sensitivity for non-SM $CP$-violation detection
using the $\ee\rt\Xi\bar{\Xi}$ process would be possible. 

\section{Summary}
The BESIII experiment has opened up a new portal in the quest to identify new, beyond-the-Standard Model
sources of $CP$ violation. However, the limits they that have now, and will set  soon, still leave two orders
of magnitude of unexplored parameter space before SM-level effects are expected. The measurement technique
that is described here is specific to the $\jpsi\rt\Xi^-\bar{\Xi}^+$ and $\Xi^0\bar{\Xi}^0$ reactions,
and the main issue impeding searches with improved sensitivities is statistics; trillion $\jpsi$~event
data samples are needed, and these could be provided by new facilities that have been proposed in China and
in Russia.

\section*{Acknowledgments}

This work is supported in part by the National Natural Science Foundation of China (NSFC) under Contracts
No. 12122509, the Korean Institute For Basic Science under project code IBS-R016-D1 and the Chinese Academy
of Science President’s International Fellowship Initiative (PIFI) program.

\bibliographystyle{ws-ijmpa}
\bibliography{BESIII-CP-refs}
\end{document}